\documentclass[11pt]{article}

\usepackage{amsmath}

\newcommand{\np}{\textsc{np}}
\newcommand{\sat}{\textsc{sat}}
\newcommand{\SAT}{\mathbf{SAT}}
\newcommand{\AND}{\wedge}
\newcommand{\OR}{\vee}
\newcommand{\ket}[1]{|#1\rangle}
\newcommand{\Or}{\mathcal{O}}
\newcommand{\ox}{\otimes}

\title{\textbf{Two Classical Queries\\
versus\\
One Quantum Query}}
\author{\textit{Wim van Dam}\\
Centre for Quantum Computation, University of Oxford\thanks{
Centre for Quantum Computation, 
Clarendon Laboratory, 
Department of Physics, 
University of Oxford,
Parks Road,
Oxford~~OX1~3PU, 
United Kingdom.
Quantum Computing and Advanced Systems Research,
C.W.I., 
P.O.\,Box~94079, 
NL--1090~GB~~Amsterdam, 
The Netherlands.
This work was supported by the European TMR Research Network 
ERP-4061PL95-1412, Hewlett-Packard, 
and the Institute for Logic, Language and Computation in Amsterdam.
Email address: \texttt{wimvdam@qubit.org}
}\\
Quantum Computing and Advanced Systems Research, C.W.I.}
\begin{document}
\maketitle
\begin{abstract}
In this note we study the power of so called query-limited computers.
We compare the strength of a classical computer that is allowed to
ask two questions to an \np-oracle with the strength
of a quantum computer that is allowed only one such query.
It is shown that any decision problem that 
requires two parallel (non-adaptive) \sat-queries on a 
classical computer can also be solved exactly by a quantum 
computer using only one \sat-oracle call, where both 
computations have polynomial time-complexity. 
Such a simulation is generally believed to be impossible for a 
one-query classical computer. 
The reduction also does not
hold if we replace the \sat-oracle by a general black-box.
This result gives therefore an example of how a quantum computer 
is probably more powerful than a classical computer. It also 
highlights the potential
differences between quantum complexity results for general oracles 
when compared to results for more structured tasks like the 
\sat-problem.
\end{abstract}
\newpage
\section{Introduction}
We consider the question if a quantum computer which is allowed to 
consult an \np-oracle (from now on a \sat-oracle) only once, 
can simulate exactly and efficiently a 
classical computer that is allowed to ask two non-adaptive 
(parallel) \sat~queries to an oracle. 
(By `non-adaptive' it is meant that the phrasing of the second
question is not allowed to depend on the outcome of the first query.)
We will see that the answer to this question is ``yes'', 
and hence for the readers familiar with the notation used for complexity 
classes:
\begin{eqnarray}
\mathbf{P}_{\mathrm{tt}}^{\mathbf{NP}[2]} & \subseteq & 
{\mathbf{EQP}}^{\mathbf{NP}[1]}~,
\end{eqnarray}
where ``tt'' denotes `truth-table reducibility' (non-adaptive calls). 
See Garey and Johnson\cite{Garay} for an introduction in 
complexity theory and some of the work by Richard 
Beigel\cite{Beigel1,Beigel2} for an overview of query-limited 
reductions.

This question is inspired by the article ``Two Queries'' by 
Harry Buhrman and Lance Fortnow\cite{Buhrman}. Its answer uses some
well known results on Deutsch's problem\cite{Deutsch}
and its one-call, exact solution\cite{Cleve}.
At the end of the note we will discuss the question if the same
result also holds for two adaptive (serial) \sat-queries.

\section{Main Result}
The classical computer is allowed to ask two questions to a 
\sat-oracle in order to solve a decision problem. 
Because the oracle calls
are non-adaptive, we can assume that there are two formulas $A$ and $B$
with which the program calculates the final, binary answer 
$F(\Or(A),\Or(B))$; where $\Or(A)$ and $\Or(B)$ are the 
respective oracle answers to the calls ``$A\in\SAT$?'' and ``$B\in\SAT$?''.
The function $F$ will therefore be of the form 
$F:{\{0,1\}}^2 \rightarrow \{0,1\}$.

The central idea is that for every possible $F$ we can transform the 
original two oracle calls $\Or(A)$ and $\Or(B)$ into a procedure
that calculates $F(\Or(A),\Or(B))$ directly, while using only one call.
For different $F$, different solutions exist for this problem. 
We start our proof by describing four of those transformations, 
after which we conclude by an exhaustive list of all possible 
functions $F$ and how they can be solved exactly by a quantum algorithm.

\subsection{Two Classical Reductions}
Classically we can combine two oracle calls $\Or(A)$ and $\Or(B)$ 
into one oracle call, in the following two ways:
\begin{eqnarray}
A\in\SAT\textrm{?}\textrm{  or  }B\in\SAT\textrm{?} 
& \Longleftrightarrow &
(A \OR B) \in \SAT\textrm{?} \\
A\in\SAT\textrm{?}\textrm{  and  }B\in\SAT\textrm{?}
& \Longleftrightarrow &
(A \AND B) \in \SAT\textrm{?}
\end{eqnarray}
This is because
\begin{eqnarray}
\exists x [A(x)] \OR \exists y[B(y)] & \Longleftrightarrow & 
\exists xy [A(x) \OR B(y)] \\
\exists x [A(x)] \AND \exists y[B(y)] & \Longleftrightarrow & 
\exists xy [A(x) \AND B(y)]~.
\end{eqnarray}
We therefore can use the equations:
\begin{eqnarray}
\Or(A) \OR \Or(B) & = & \Or(A\OR B) \\
\Or(A) \AND \Or(B) & = & \Or(A\AND B) 
\end{eqnarray}
where the left-hand ``$\OR$'' and ``$\AND$'' are interpreted as 
binary functions.

\subsection{Two Quantum Reductions}
In the quantum case we can use the one-call solution to 
Deutsch's problem\cite{Cleve}. That is, we start with the system in 
the state:
\begin{eqnarray} \label{eq:dj1.1}
\ket{\textrm{begin}} & = & 
\frac{1}{2}
(\ket{A}+\ket{B})~\ox~(\ket{0}-\ket{1})~,
\end{eqnarray}
and write down the respective oracle values $\Or(A)$ and 
$\Or(B)$ in the rightmost qubit. This yields:
\begin{eqnarray}
\pm\frac{1}{2}(\ket{A}+\ket{B})~\ox~(\ket{0}-\ket{1}) & \textrm{ if } & 
\Or(A)=\Or(B)\\
\pm\frac{1}{2}(\ket{A}-\ket{B})~\ox~(\ket{0}-\ket{1}) & \textrm{ if } & 
\Or(A) \neq \Or(B)~.
\end{eqnarray}
By applying the unitary mapping 
\begin{eqnarray}
\ket{A}  \rightarrow  \frac{1}{\sqrt{2}}(\ket{0}+\ket{1}) 
& \textrm{ and }& 
\ket{B}  \rightarrow  \frac{1}{\sqrt{2}}(\ket{0}-\ket{1}) 
\end{eqnarray}
to the first register,
we thus become for the final state:
\begin{eqnarray} \label{eq:dj1.2}
\ket{\textrm{end}} & = & 
\pm\frac{1}{\sqrt{2}}
\ket{\Or(A)\oplus \Or(B)}~\ox~(\ket{0}-\ket{1}).
\end{eqnarray}
It is therefore that in the quantum case, we can melt the two oracle calls
of the expression
\begin{eqnarray} \label{eq:dj1.3}
&
(A\in\SAT\AND B\not\in\SAT) \OR (A\not\in\SAT\AND B\in\SAT)\textrm{?}~, 
&
\end{eqnarray}
into one quantum oracle call.

Another variant of this one-call trick can be employed if the 
beginning state is of the form:
\begin{eqnarray} \label{eq:dj2.1}
\ket{\textrm{begin}'} & = & 
\frac{1}{2}
(\ket{A}+\ket{A\AND B})~\ox~(\ket{0}-\ket{1})~,
\end{eqnarray}
such that the ending state will be:
\begin{eqnarray} \label{eq:dj2.2}
\ket{\textrm{end}'} & = & 
\pm\frac{1}{\sqrt{2}}
\ket{\Or(A)\oplus \Or(A\AND B)}~\ox~(\ket{0}-\ket{1})~.
\end{eqnarray}
This evaluation corresponds to the two-call expression
\begin{eqnarray} \label{eq:dj2.3}
& 
(A\in\SAT) \AND (B\not\in\SAT)\textrm{?} 
&
\end{eqnarray}
In the next section we will show how the above four reductions can be 
used to calculate all possible decision functions 
$F(\Or(A),\Or(B))$ with only one quantum oracle call.

\subsection{Using the Four Reductions}
We define the accepting set by $S=\{(a,b)|F(a,b)=1\}$, where $a$ and $b$
range over the possible values of $\Or(A)$ and $\Or(B)$, or simply:
$(a,b) \in \{0,1\}^2$. The set $S$ is therefore a subset of 
$\{(0,0),(0,1),(1,0),(1,1)\}$. Now we show how for every possible
$S$, we can construct a quantum oracle call that answers the decision
question ``$(\Or(A),\Or(B))\in S$?''.

Without loss of generality we assume that $|S|\leq 2$. Hence, we can 
distinguish the following 3 cases.
\begin{itemize}
\item{$|S|=0$: This is trivial. The protocol always answers ``$0$''.}
\item{$|S|=1$: There are 3 sub-cases here: 
\begin{itemize}
\item{$S=\{(0,0)\}$: Use the classical oracle call $\Or(A\OR B)$.}
\item{$S=\{(1,1)\}$: Use the classical oracle call $\Or(A\AND B)$.}
\item{$S=\{(0,1)\}$ or $S=\{(1,0)\}$: This case is calculated by the second
version of the one-call Deutsch solution, as explained in the Equations
\ref{eq:dj2.1}, \ref{eq:dj2.2}, and \ref{eq:dj2.3}.}
\end{itemize}}
\item{$|S|=2$: This possibility has essentially 2 sub-cases:
\begin{itemize}
\item{$S=\{(0,0),(1,0)\}$,
$S=\{(0,1),(1,1)\}$,
$S=\{(0,0),(0,1)\}$, or
$S=\{(1,0),(1,1)\}$: All these subsets only depend on one of the classical
oracle values and can therefore be solved by one oracle call of the
form $\Or(A)$ or $\Or(B)$. }
\item{$S=\{(0,0),(1,1)\}$ or $S=\{(0,1),(1,0)\}$: This case can be 
recognised by the original one-call solution of Deutsch's problem, as shown
in the Equations \ref{eq:dj1.1} till \ref{eq:dj1.3}.}
\end{itemize}}
\end{itemize}
We thus see how a quantum computer can solve all possible decision 
problems described by $S$. This implies that any decision problem
that can be solved efficiently on a classical computer with the use 
of two non-adaptive \sat-queries, can also be solved 
by a quantum computer which uses only one \sat-query (again with
polynomial time complexity). For a classical computer with one query
at its disposal this is generally believed to be 
impossible\cite{Beigel1,Beigel2,Buhrman}.

\section{Conclusion, Question and Reminder}
We have shown how a quantum computer with one query to an \np-oracle
can solve efficiently all decision problems that a classical computer 
can calculate in polynomial time with the help of two non-adaptive 
\np-oracle calls.

It is not obvious how to generalise this result to the case of two
\emph{adaptive queries.\/}
This is the scenario when the input for the second oracle call can depend
on the outcome of the first oracle call. 
It is already known\cite{Beigel1} that this complexity class 
is equivalent with the classical case with three non-adaptive \np-queries.
The question thus becomes:
\begin{eqnarray}
\mathbf{P}^{\mathbf{NP}[2]}
=
\mathbf{P}^{\mathbf{NP}[3]}_{\mathrm{tt}} & 
\stackrel{\mathrm{?}}{\subseteq} & {\mathbf{EQP}}^{\mathbf{NP}[1]}~.
\end{eqnarray}
A result by Beals \emph{et al.\/}\cite{Beals} showed that for 
\emph{oracles without any structure,\/} the full $N$ calls are required 
to exactly calculate the \textsc{and} function over $N$ black-box values. 
If we apply this result to the 
$N=2$ case (that is, the problem ``$\Or(A)\AND\Or(B)?''$), 
then we see that the \np-structure of the oracle 
is an essential part in the above proof.
The main result of this note is therefore also a reminder that the 
lower bounds for \emph{general\/} black-boxes do not tell us everything 
there is to know about the potential quantum speed-up for a 
\emph{specific\/} computational problems.

\section*{Acknowledgements}
I would like to thank Harry Buhrman for useful conversations about 
several aspects of structural complexity theory.

\end{document}